# Superconductivity in hole-doped germanium point contacts

N. V. Gamayunova[1], M. Kuzmiak[2], P. Szabó[2], P. Samuely[2], and Yu.G. Naidyuk[1]

[1]*B. Verkin Institute for Low Temperature Physics and Engineering of the National Academy of Sciences of Ukraine 47 Nauky Ave., Kharkiv 61103, Ukraine*

[2]*Centre of Low Temperature Physics, Institute of Experimental Physics, Slovak Academy of Sciences Watsonova 47, SK-04001 Košice, Slovakia*

E-mail: naidyuk@ilt.kharkov.ua

**Abstract**

We have observed superconductivity in heavy p-doped Ge by measuring of differential resistance *dV/dI(V)* of Ge – PtIr point contacts. The superconducting (SC) features disappear above 6 K or above 1T, what can be taken as the critical temperature and the critical magnetic field, respectively. The observed *dV/dI(V)* spectrum with Andreev reflection like features was fitted within one-gap Blonder-Tinkham-Klapwijk model. The extracted SC gap demonstrates Bardeen-Cooper-Schrieffer-like behavior with $2\Delta/k_BT_c = 10\pm1$ ratio, which is much higher that expected for conventional superconductors. Magnetic field suppresses Andreev reflection features, but the SC gap moderately decreases in magnetic field similarly as it was observed previously for the type–II superconductors, including nickel borocarbide and iron-based superconductors. Curiously, we have not yet observed superconductivity in n-doped Ge with a similar dopant concentration.

**Introduction**

Since the prediction of possibility of superconductivity in doped semiconductors such as silicon and germanium by theoretical calculations in 1964 [1], many studies have been performed to find superconductivity in such materials. Superconductivity in the metallic high-pressure Si and Ge phases was discovered in 1966 and demonstrated the temperatures of superconducting transition $T_c$=5.35 K for Ge at 11,5 GPa and $T_c$=6.7 K for Si at 12 GPa [2]. Higher $T_c$ = 8.2 K was found in metallic Si phases at high-pressure of 15 GPa [3], however decrease of $T_c$ up to 3.5 K was observed in Si by pressure increase up to 45 GPa [4]. To reach superconductivity at ambient pressure conditions was enough difficult task, because it is needed heavy doping beyond the metal–insulator transition for the carrier density in these materials to be sufficient to induce a superconducting state at low temperatures.

Ge, in a comparison with Si, looks less promising for creating superconductivity in terms of the standard Bardeen-Cooper-Schrieffer (BCS) model because of the lower Debye temperature 360 K versus 640 K in Si [5]. For the formation of a superconducting state in Ge, it turned out that an acceptor concentration higher than the equilibrium solubility in a solid is needed. This requires complex preparation techniques like a high-pressure high-temperature synthesis, the chemical vapor deposition, gas immersion laser doping and high-fluence ion implantation combined with rapid thermal annealing, or flash lamp annealing [6]. The theoretical studies predicted a weak superconductivity in heavily hole-doped Si and Ge [7], and its observation at ambient pressure revealed $T_c$ = 0.45 K in Al- and Ga-doped Ge [6], 0.5 K [8] and 1.4 K [5] in Ga-doped Ge.

A few decades ago, degenerated Ge was studied by Yanson point-contact (PC) spectrosco-py [9] to search for electron-phonon interaction function, which is responsible for superconducti-vity in the BCS theory. Some features associated with the electron-phonon interaction were registered [10] (see Fig.S1 in Supplement). Here we continue the PC study of degenerated Ge with various concentration of charge carriers. Searching for superconductivity in this material was reactivated after reporting discovery the superconductivity above 10K in silicon by PCs technique [11]. As a result, we observed superconductivity in PCs based on p-type Ge. Moreover, Andreev reflection-like features were registered, which allowed us extract the superconducting gap and its temperature and magnetic field dependence.

**Method**

We performed the PC spectroscopic measurements of heavy p-doped Ge crystals, where the concentration range of gallium impurities was $2·10^{17}$–$2·10^{18}$ см$^{-3}$ [12]. The PCs on the base of Ge samples were created using the "needle-anvil" technique [9] by touching a sharpened Pt80/Ir20 wire with the diameter of 0.25 mm to the planes of the Ge crystals at helium temperature. Before that, the surfaces of Ge crystals were polished by sandpaper and cleaned in acetone at a room temperature. The first derivatives *dV/dI(V)* of current-voltage characteristics of germanium PCs were measured in the temperature range 1.5-10 K and magnetic field up to 2 T. The differential resistance spectra of PCs *dV/dI(V)* were recorded by sweeping the *dc* current *I* on which a small *ac* current *i* was superimposed using a standard lock-in technique [9].

**Results and discussion**

Typical *dV/dI(V)* spectra for all samples show overall "semiconducting" behavior (that is *dV/dI(V)* decreases with a voltage) with a zero-bias maximum (see Fig. 1(a)). Occasionally, repeatedly pressing the PC needle to the sample plane caused opening of additional spectral features at around the zero-bias (see Fig. 1(b) and Fig. S2 in Supplemet).

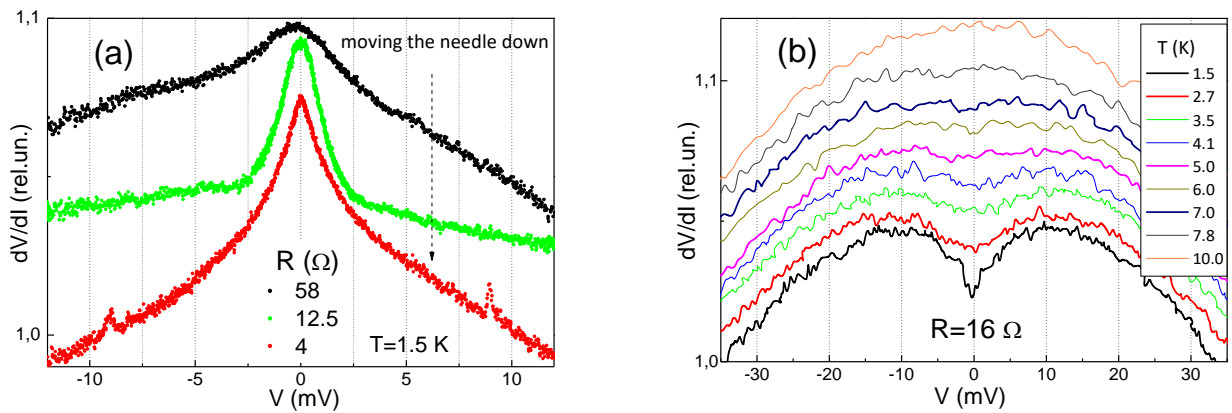

Fig. 1. (a) *dV/dI(V)* spectra of p-type Ge (p=10$^{18}$ см$^{-3}$) – PtIr PCs at T=1.5K. The needle moves down to the bulk material of the sample that results in decrease of the PC resistance. (b) Temperature series of normalized *dV/dI(V)* spectra of p-type Ge (p=2.8·10$^{17}$ см$^{-3}$) – PtIr contact.

Fig. 1 shows the *dV/dI(V)* spectra of PC on the base of Ge with carrier concentration of 2.8*10$^{17}$ cm$^{-3}$ with the zero-bias minimum at 1.5K. This minimum reduces its intensity with temperature increase and disappears at about 5-6 K. This feature of the temperature series is inherent to originally superconducting materials with a critical temperature of about 6 K.

Fig. 2 demonstrates *dV/dI(V)* spectra of PC on the base of Ge with carrier concentration of 10$^{18}$ cm$^{-3}$ with the Andreev-like double minimum structure, where minima are located of about +/–3 mV at low temperature. Similar to the previous case, these minima are suppressed with temperature or magnetic field increase and finally they disappear above 6K or 1T. The appearance of such features may be caused by the combination of general "semiconducting" *dV/dI(V)* behavior and superconducting features around zero-bias. This gave us the opportunity to use these features for determining of the superconducting gap value in hole-doped germanium within standard Blonder-Tinkham-Klapwijk (BTK) approach [13].

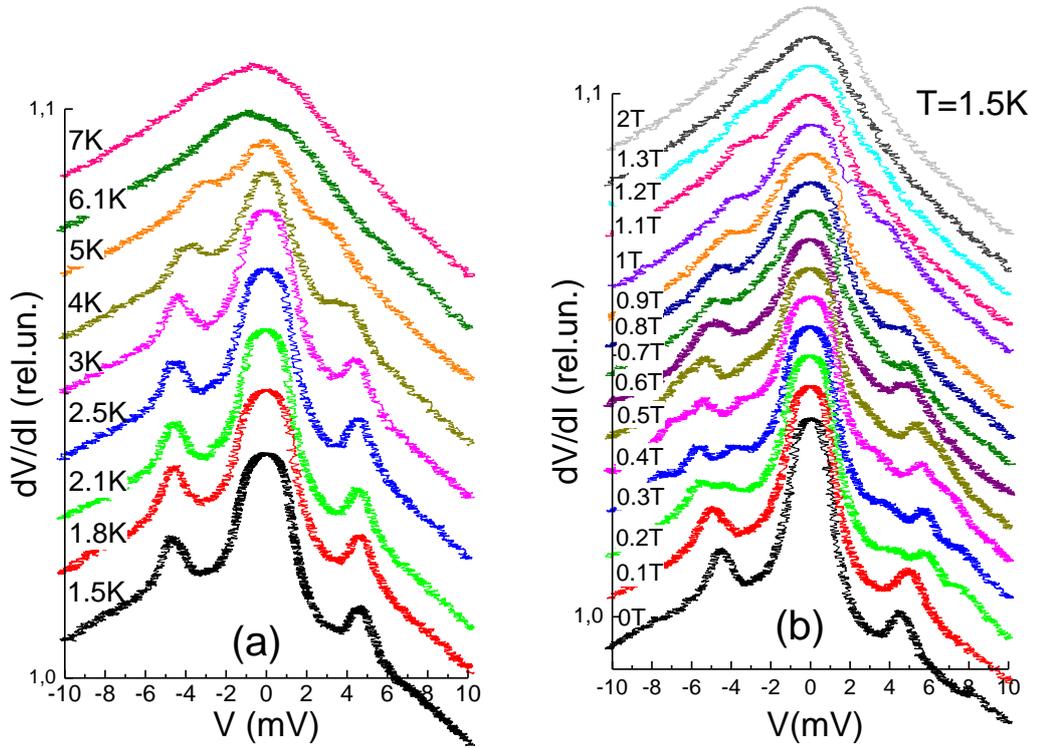

Fig. 2. Temperature (a) and magnetic field (b) series of experimental *dV/dI(V)* spectra of p-type Ge(p=$10^{18}$ см$^{-3}$) – PtIr contact with R(0)=16 Ω. *dV/dI(V)* spectra are shifted vertically for a better overview.

Before the fit procedure the measured Andreev-like *dV/dI(V)* spectra were symmetrized and normalized on the curve in the normal state at 7 K in the case of temperature series and curve at 7 K or at 2 T in the case of magnetic field series. Then these curves were fitted (Fig. 3) by one-gap BTK model [13], which describes the Andreev reflection at the boundary between a normal metal and a superconductor in view of such parameters as Δ (the superconducting gap), Z (the contact barrier strength) and Γ (the spectral smearing parameter). The calculated theoretical curves correspond reasonably well to the experimental ones (see Fig. 3) excluding the side maxima around 5 mV, which arise from a non-Andreev-reflection contribution to the *dV/dI(V)* spectra [14].

The results of the temperature and magnetic field dependences of the gap and extra fit parameters are presented in Fig. 4. The temperature gap behavior (see Fig. 4 (a)) is close to the BCS-like curve. Obtained from the BCS extrapolation the gap value Δ=2.4 meV[1] leads to the ratio $2\Delta/k_BT_c$=10±1 if $T_c$=5–6 K, which is surprisingly a much higher value, then for conventional superconductors. By the way, this ratio for doped Si is also high (about 7) as it follows from the data in Fig. 3(e) [12]. There are also other cases of extremely large value of $2\Delta/k_BT_c$. As a recent example – the mesoscopic point contacts between pure silver and the 3D Dirac semimetal $Cd_3As_2$ demonstrate the gap magnitude 6.5 meV and $\Delta/k_BT_c = 10$ (if the low $T_c \approx 6$ K is taken) along with the pseudogap features up to 13K [15].

---

[1] We have obtained similar value of Δ=2.2 meV for another PC with larger resistance of 43 Ω (see Fig. S3 in Supplement).

As can be seen from Fig. 2 (b), magnetic field gradually suppresses superconducting features until they disappeared around $B_c \approx 1.3$T. At the same time, the superconducting gap (Fig. 4(b)) does not show strong decrease approaching $B_c$. Similar behavior was observed for nickel borocarbide [16, 17] and iron-based superconductors [18], where such gap behavior versus magnetic field was supposed to have relation to a multiband scenario, electronic DOS modifications in the mixed state and/or vortex pinning/distribution near the contact interface. However, such magnetic field gap behavior is still not completely understood. The latter together with high $2\Delta/k_BT_c$ ratio may indicate some peculiarities of superconducting state in doped Ge. Although E. Bustarret in the survey "*Superconductivity in doped semiconductors*" [19] summarized that **"**superconducting semiconductors hint at a conventional pairing mechanism …. However, quantitative agreement with simple models was not yet demonstrated over a wide doping range for any of the "conventional" systems, while additional evidence is still required to assess the status and potential of *germanium* and silicon carbide".

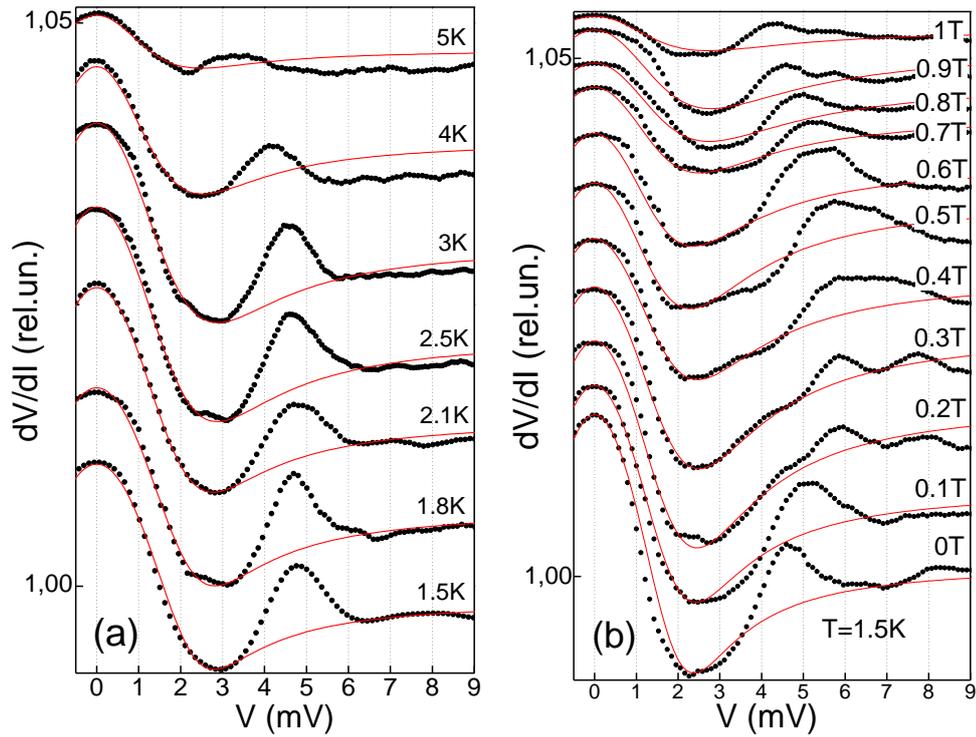

Fig. 3. Theoretical fit (solid lines) for the temperature (a) and magnetic field (b) series of normalized $dV/dI(V)$ spectra (dots) from Fig. 2 within single-gap BTK model.

The low intensity of the spectra should be noted, which is only of about few percent (see Fig. 3). Therefore, when fitting data, we included the scaling parameter $S=dV/dI(V)_{exp}/dV/dI(V)_{theor}$, which reflects (see Fig. 4 (c, d)) that the experimental curve intensity is about one order of magnitude smaller than that of theoretical one. Most likely, low intensity is connected with a small volume of the superconducting phase in the PC. Other reasons of a low S are discussed in Appendix of Ref. [16] (or in Supplement). Also, clear rise of the spectral smearing parameter $\Gamma$ is seen in a magnetic field. Such $\Gamma$ behavior was observed for type-II superconductors [16-18, 20, 21] and is discussed considering pair-breaking in magnetic field and vortex mixed state [20, 21].

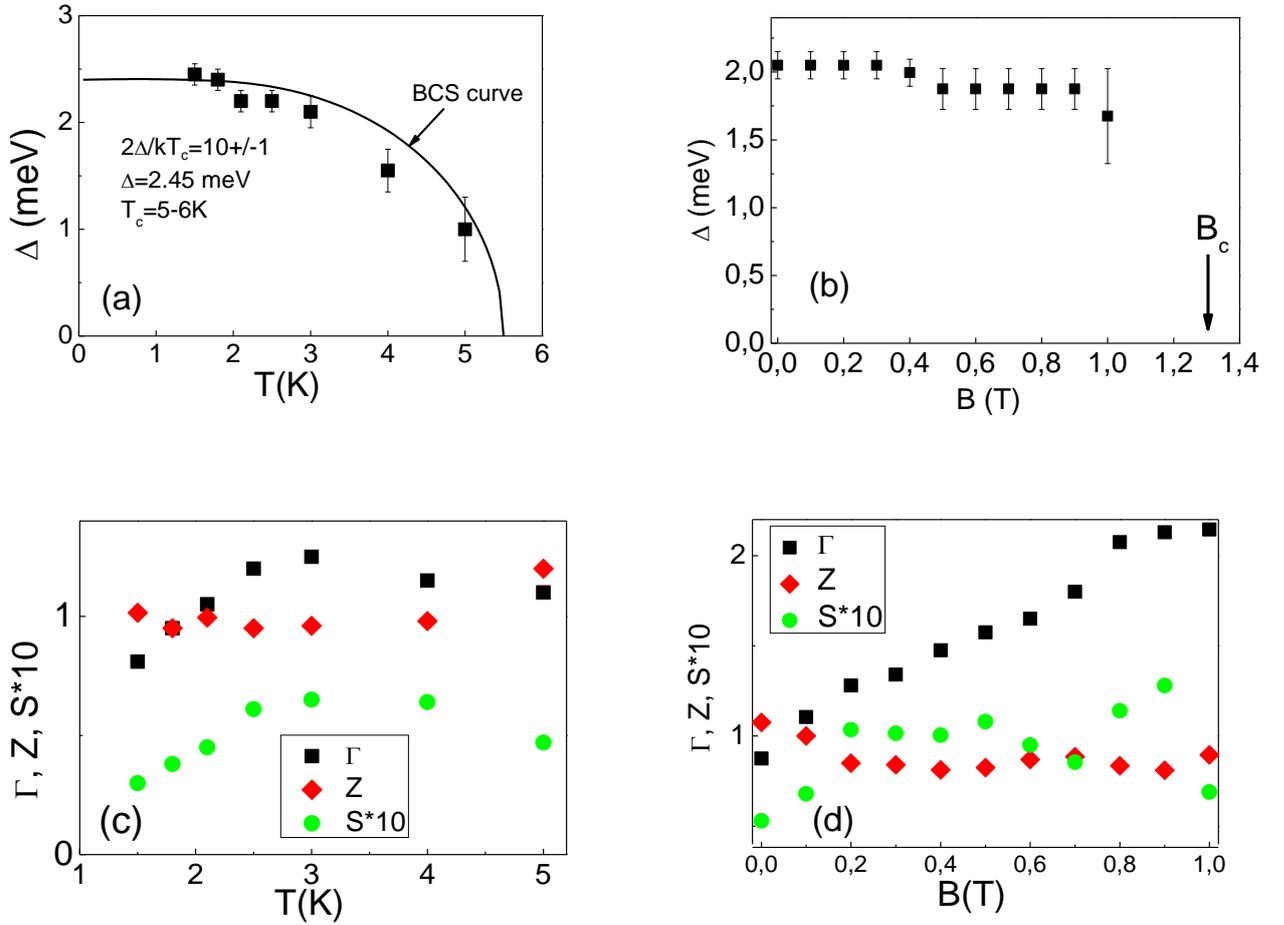

Fig. 4. Calculated temperature (a, c) and magnetic field (b, d) dependences of the fitting parameters: supercoducting gap $\Delta$, broadening parameter $\Gamma$, barrier Z and scaling parameter S for spectra shown in Fig.3.

As to the nature of observed superconductivity in Ge PCs. Considering, that the superconducting phase with $T_c$=5.35 K in Ge appears under pressure at 11.5 GPa [2], it is very likely that the local pressure in PC is the reason of observed superconductivity with similar $T_c$. In addition, recent first-principle calculations predicted [22], that inducing and tuning superconducting states in semiconductors is possible due to shear strains. The latter can also take place by pressure type PCs.

Finally, we have not yet observed superconductivity in n-doped Ge with similar dopant concentration.

**Summary**

We have observed superconductivity induced in heavy p-doped Ge by PCs. The emergence of superconductivity is likely connected with pressure and/or strain effects by mechanical type formation of PC. The superconducting features disappear above 6 K or above 1T, what can be taken as the critical temperature and magnetic field of superconducting transition, respectively. The *dV/dI(V)* spectra have been reasonably fitted with one-gap BTK model. Extracted SC gap demonstrates BCS-like behavior with $2\Delta/k_BT_c$= 10±1, which is 3 times higher than the BCS ration for conventional superconductors. Magnetic field suppresses SC features, but the gap moderately decreases in magnetic field similarly as it was observed for borocarbide [16, 17] and iron-based superconductors [18].

**Acknowledgments**

The studies were conducted as part of a joint project between National Academy of Scien-ces of Ukraine and Slovak Academy of Sciences. NVG would like to thank the Institute of Experimental Physics (Košice) for their hospitality and O.E. Kvitnitskaya and L.V. Tyutrina for useful discussions. The work in Slovakia was supported by the projects APVV-18-0358, VEGA 2/0058/20, VEGA 1/0743/19 and by U.S. Steel Košice.

**Supplement**

1) Previous PC measurements with Ge

Main panel of Fig. S1 shows our data from Ref. [10]. Insert shows *dV/dI* for corresponding curves from the main panel after their integration. We see that *dV/dI* demonstrate minimum at zero bias, which is more likely connected with superconductivity, but not with a peculiar electron-phonon interaction as it was supposed in Ref. [10].

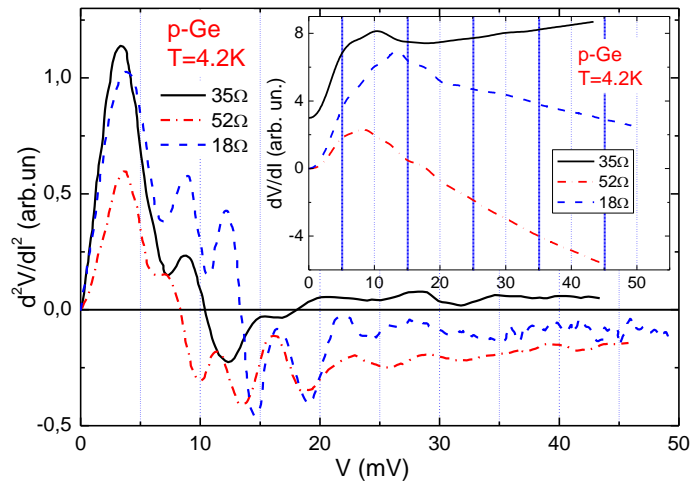

Fig. S1. Second derivative $d^2V/dI^2$ spectra of three homocontacts (!) of p-Ge (p=$10^{18}$ см$^{-3}$) with different resistance from Ref. [10]. Inset shows result of integration of the corresponding curves from the main panel that it their first derivatives.

2) Additional *dV/dI(V)* data on germanium PCs

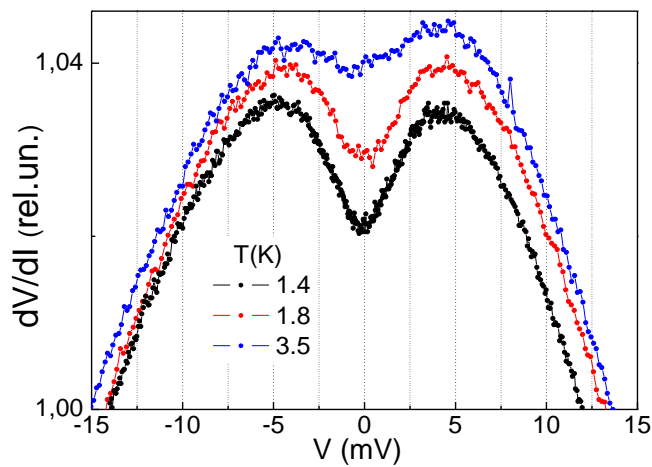

Fig. S2. The incomplete temperature series of *dV/dI(V)* spectra of Ge(p=2.8·$10^{17}$ см$^{-3}$) – PtIr point contact which was destroyed at the temperatures above 3.5 K. *dV/dI(V)* spectra are shifted vertically for a better overview. The distinct zero-bias minimum on overall "semiconducting" behavior of the curves refers to the superconducting state in the Ge sample, which is suppressed with the temperature increase.

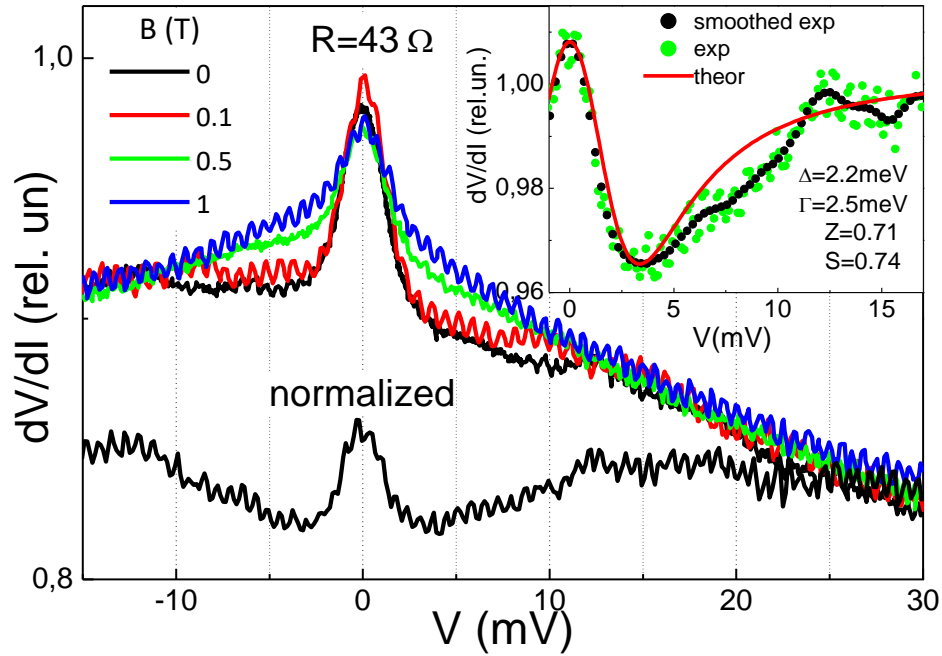

Fig. S3. *dV/dI* spectra of p-Ge(p=10$^{18}$ см$^{-3}$) – PtIr PC in magnetic field at 1.5K. Bottom curve is normalized zero field *dV/dI,* that is *(dV/dI)*$_{0T}$ */(dV/dI)*$_{2T}$. This curve is shifted down by 0.15. Inset: Theoretical fit (red line) for the normilized curve from the main panel (dots) within single-gap BTK model. Obtained superconducting gap value Δ=2.2 meV is close to the gap value, which is extracted from the fit on Fig.4.

3) <u>Reasons for the low value of the scaling parameter (following the Appendix of Ref. [17])</u>

We included in the fit the ratio of intensity of the calculated and the experimental curves marked as S. For instance, *S* = 1 means that the calculated curve fits the measured *dV/dI* also in absolute values. If *S* < 1, the measured *dV/dI* has a reduced intensity due to suppression of the Andreev reflection signal or due to an opening of some non-AR channel in the PC conductivity. E.g., *S* < 1 can occur when the Fermi surface is only partially gapped in the SC state. The more common explanation of non-SC regions in the PC is multicontact scenario including PCs with suppressed superconductivity, which shunt the main signal. Also, a contribution to the conductance from the normal vortex cores in the superconductor was taken into account in Ref. [21] to describe a progressive suppression of the AR features in an external magnetic field.